\documentstyle[11pt,english,epsfig]{article}
\begin{document}
{\LARGE\bf
\begin{center}
              Effect of Zero Modes on the Bound-State Spectrum \\
                 in Light-Cone Quantisation   
\end{center}}
\vspace*{1.0cm}
   {\begin{center} 
{\large\sc Annette S. M\"uller}\\
Sektion Physik, 
Ludwig--Maximilians--Universit\"at M\"unchen\\ 
D--80333 M\"unchen\\
[0.5cm]
{\large\sc  Alexander C. Kalloniatis}\\
Institut f\"ur Theoretische Physik III, 
Universit\"at Erlangen--N\"urnberg\\
D--91058 Erlangen\\
[0.5cm]
{\large\sc Hans--Christian Pauli}\\
Max--Planck--Institut f\"ur Kernphysik\\
D--69029 Heidelberg\\
[0.7cm] 
25 March 1998\\
[2.7cm] { }
\end{center}}

Preprint Number: FAU TP3-98/7.

\begin{abstract}
We study the role of bosonic zero modes in light-cone quantisation
on the invariant mass spectrum for the simplified setting
of two-dimensional SU(2) Yang-Mills theory coupled to massive
scalar adjoint matter. Specifically, we use
discretised light-cone quantisation where the momentum
modes become discrete. Two types of zero momentum mode
appear -- constrained and dynamical zero modes.
In fact only the latter type of modes turn out to  
mix with the Fock vacuum. Omission of the constrained 
modes leads to the dynamical zero modes being controlled
by an infinite square-well potential. We find that
taking into account the wavefunctions for these modes
in the computation of the full bound state spectrum of the two
dimensional theory leads to 21\% shifts in the masses
of the lowest lying states.  

\end{abstract} 
\newpage

\section{Introduction}

It has always been clear that a `solution' of quantum chromodynamics  
is constituted not just by numerical reproduction of the hadron spectrum
to a certain accuracy, but also an accompanying
intuitive picture of what are the essential degrees
of freedom involved in the building up of hadrons.
In other words, what are the `quasiparticles'? 
On the one hand, these quasiparticles must have
something to do with the constituent quarks which
are useful for, at the very least, classifying
the hadron spectrum. On the other hand, they should
emerge from some systematic treatment of the QCD
Lagrangian. Finding adequate field theoretic
and nonperturbative formalisms in which this can take
place has been an ongoing problem for many years.

One must recognize that lattice gauge theory has 
taken some major steps in this direction in recent
years: it has gone beyond mere `simulation' of the
hadronic spectrum (see for example, \cite{Lat})
to quite refined calculations attempting to focus
on specific `physical mechanisms' related, for example, to confinement
or symmetry breaking (as in, say, \cite{Mon}). 
However, the focus on the putative `essential'
degrees of freedom is something that has to be imposed by the theoretician on
the calculation after generation of a large number
of arbitrary field configurations, rather than emerging   
naturally within the lattice gauge formalism.
The complementary formalism of light-cone, or more properly,
Dirac front-form \cite{Dir49}, dynamics offers a framework in which the
`quasiparticles' naturally are central from the beginning
\cite{Wei66,LeB80,PaB85,PHW90}.
The reason for this is, by now well-known and even  
somewhat clich{\'e}d if not even completely correct,
the so-called simplicity of the light-cone vacuum \cite{Wei66}.
The essence of this simplicity is that the light-cone
momentum operator is positive-definite {\it in the absence
of zero momentum modes} and thus no states built out of
Fock operators on the Fock vacuum can mix with this vacuum.
In that case then, we have the ideal candidate for 
the vacuum on top of which quasiparticles -- constituent
quarks -- can be built as elementary excitations. 

However zero momentum modes of the basic fields in the
Lagrangian are in principle present and cannot be {\it ad hoc} 
thrown away \cite{MaY76}.
An open question which we will address in this
paper is: can they have in fact any effect
on the mass spectrum of a theory in the continuum limit?
 
The `zero mode problem' of light-cone quantisation 
(see, for example, \cite{HKW91,Rob93}) is
the problem of how to now introduce the zero momentum
modes without simultaneously
destroying the above pleasant picture. The formalism of
discretised light-cone quantisation (DLCQ)
developed originally in \cite{PaB85} ironically
in the absence of zero modes, 
compactifies the light-cone `spatial' direction 
$x^-= (x^0 - x^1)/\sqrt{2}$
leading to discrete longitudinal momenta.
It thus offers an elegant framework for studying this problem because
the zero modes can be cleanly separated out. Despite the
eminently unphysical nature of imposing periodic boundary
conditions on points on a null-plane \cite{LTYL91} -- and the
concomitant loss of some physics which can take place 
\cite{Rob96} -- this approach has
enabled some progress within various gauge theories (for a review,
see \cite{BPP97}) and is finding 
application now in string theories and M-theory \cite{BFSS97,BS97}. 

In DLCQ, some zero modes indeed appear in the
problem as genuine degrees of freedom, with conjugate
momenta \cite{KPP94}. 
These modes can indeed mix with the vacuum
but because they do not have any transverse momentum dependence
(see, for example, \cite{KaR94})  
they do not obscure the quasiparticle interpretation of the
field Fock modes.  But they can carry other internal quantum numbers 
and thus are not so undesirable: by mixing between these modes
and the Fock vacuum, the true vacuum can acquire nontrivial
quantum numbers thus breaking a symmetry spontaneously.
Moreover, this can happen without destroying the 
physical picture of quarks as quasiparticles. 
Other zero modes in DLCQ however have no conjugate momentum
and are constrained \cite{MaY76,FNP81}. 
They are -- in other words --
dependent on the true dynamical degrees of freedom,
including, in non-Abelian gauge theories,
{\it the dynamical zero modes}. This dependence is given through
constraint equations arising from the equations of motion
which in general are highly nonlinear. Thus, though in
DLCQ we have managed to preserve the simple physical picture
sought at the outset, we have arrived at a mathematically
difficult problem: solving these constraint equations.
Pioneering work in this direction was done for two-dimensional
scalar theory in \cite{scalar}.   

The degree of difficulty of this problem in a gauge theory 
has warranted some of us studying its analogue
in the simpler setting of two-dimensional
SU(2) Yang-Mills theory coupled to scalar adjoint matter
fields -- originally studied by \cite{DKB93}. 
In this model, the problem remains significantly complicated.
But the following insights have, to date, been attained
through the works \cite{PKP95,PK96,Kal96}.
The dynamical zero modes are here --- as is expected in fact
for $3+1$ dimensional gauge theories as well -- strictly
time dependent fields which are better expressed by  
Schr{\"o}dinger wavefunctions rather than a Fock expansion.
These wavefunctions are eigenfunctions of the zero mode
projected Hamiltonian, the essential dynamics expressed
in the effective potential for these variables. Neglecting
the constrained zero mode contributions, the effective
potential turns out to be an infinite square-well whose wavefunctions
can be exactly determined \cite{PKP95}. Solving the constrained
zero mode within certain approximations appears to 
generate a centrifugal barrier but itself of
a height insufficient to significantly cause mixing of
symmetric and antisymmetric states \cite{Kal96}. 
It is not yet clear if the small barrier height is a consequence
of the severity of the approximations used in \cite{Kal96}.
In any case, a square well potential is at least a good starting point  
for examining the more interesting question:
if zero modes can ever in practice affect the invariant mass spectrum 
of bound states. This we carry out in this letter. 

In the absence of zero modes the spectrum for two-dimensional
gauge theory with scalar adjoint matter was computed in 
\cite{DKB93}. A similar spectrum is obtained
by ``freezing'' the dynamical zero mode, as shown in \cite{PaB96}.  
The specific question we are interested in answering is whether
inclusion of zero modes can lead to {\it shifts} in the invariant
masses of the low energy spectrum
computed in these two works. In the present paper, we provide
numerical evidence that in fact this is the case. 
We show that, due to the dynamical zero mode, the lowest energy
level in the SU(2) theory is shifted upwards by 
twenty one percent after extrapolation to the continuum.

In the following  we briefly review the formalism for two-dimensional, 
SU(2) Yang-Mills theory coupled to scalar matter. 
We then derive the invariant mass
eigenvalue problem in the presence of the zero mode
with a two particle truncation on the Hilbert space followed by 
the results for the
low energy spectrum from the numerical solution of this problem. 
We summarise our result finally and comment on its possible significance.

\section{Review of Two-Dimensional Model}                                       
                                                                                
The formalism of the model we shall consider
has been extensively reviewed in the two previous
papers \cite{PKP95,Kal96}. Because we wish to write down
the Hamiltonian in as short a space as possible
we shall here omit much of the detail of the quantisation
and formalism.
Our light-cone conventions                                                      
will be those of \cite{KoS70}:                                                  
$ x^\pm \equiv (x^0 \pm  x^1)/\sqrt{2}$.                                        
The theory we consider is (1+1) dimensional                                     
non-Abelian gauge theory covariantly coupled to massive scalar adjoint matter   
\begin {equation}                                                               
    {\cal L}  = {\rm Tr} \Bigl(                                                 
        - {1\over2} {\bf F}^{\alpha\beta} {\bf F}_{\alpha \beta}                
       + {\bf D}^\alpha \Phi   {\bf D}_\alpha \Phi                              
        - \mu_0^2 \Phi^2 \Bigr)                                                 
       \ .                                                                      
\end   {equation}                                                               
This theory can be regarded as pure SU(2)                                       
Yang-Mills in $2+1$ dimensions dimensionally reduced to $1+1$.
The field strength tensor and                                                   
covariant derivative $ {\bf D}_\alpha$ are respectively defined                 
by                                                                              
${\bf D}^\alpha = \partial^\alpha + i g [{\bf A}^\alpha, \cdot ] $
and $ {\bf F}^{\alpha \beta} = \partial^\alpha {\bf A}^\beta                    
  - \partial^\beta {\bf A}^\alpha + i g [{\bf A}^\alpha,                        
{\bf A}^\beta ]$.  
As in \cite{PKP95} we represent field matrices                                  
in a colour helicity basis                                                     
$\Phi = \tau ^3 \varphi_3 + \tau ^+ \varphi_+ + \tau ^- \varphi_-$
with $\tau^a=\sigma^a/2$.               
We use the light-cone Coulomb gauge                                             
$ \partial_-  {\bf A}^+ = 0$, which preserves the zero mode                     
of ${\bf A^+}$.                                                                 
Then a single rotation in colour space                                          
suffices to diagonalise the SU(2) colour matrix 
$ {\bf A}^+ = A^+_3 \tau_3$. 
The quantum mode $ A^+_3 $ has a conjugate momentum                             
$  p \equiv \delta L / \delta  (A^+_3)  = 2 L \partial_+ A^+_3  $               
and satisfies the commutation relation                                          
$ \bigl[ A^+_3 , p \bigr] =                                                 
 \bigl[ A^+_3 , \ 2 L \partial_+ A^+_3 \bigr] = i $ .                    
In the following it will be useful to invoke the dimensionless                  
combination                                                                     
\begin {equation}                                                               
   z \equiv {{g  A^+_3  L} \over \pi} \ .                                       
\end {equation}                                                                 
There are additional global symmetries which can be seen in                     
terms of this mode. First,                                                      
Gribov copies \cite{Gri78,Sin78} correspond to                                  
shifts $ z \rightarrow  z + 2n, n \in Z$.                                       
Shifts $z \rightarrow z + (2n + 1), n\in Z$ are `copies' generated by           
the group of centre conjugations of SU(2).                                      
The finite interval $0 <  z  < 1$ is called the                                 
fundamental modular domain, see for example \cite{vBa92}.                       
Switching to the variable $\zeta = (z - \frac{1}{2})$
enables realisation of the symmetry 
$z \rightarrow - z$ and the composite symmetry 
$z \rightarrow (1 - z)$ in a convenient way. 
After restriction to the fundamental               
domain, one is left with symmetry under reflection about 
$z=\frac{1}{2}$ or $\zeta=0$.                                         
We shall work in Schr\"odinger representation                                   
for this quantum mechanical degree of freedom,
\begin{equation}                                                                
\Psi_r[\zeta] = \langle \zeta| r\rangle                               
\ .
\end{equation}
It is the effect of these wavefunctions in the computation
of the full spectrum of the theory that we shall be
concerned with.

The diagonal component of the hermitian scalar field $\Phi$                  
is $ \varphi_3$. This is the field in which the constrained
zero mode appears, studied in \cite{Kal96}. As mentioned in the introduction,
we omit the consequences of the constrained mode on the
spectrum, and hence
drop it in the following. Then, the following field expansion can
be given for this field
\begin {equation}                                                               
  \varphi_3 (x^-) =                                       
    {1 \over \sqrt{4\pi}} \sum_{l =1}^{\infty} \Bigl(                          
  \  a_l \  w_l \ {\rm e}^{-i l   {\pi\over L}   x^-}                           
  + \  a^{\dag}_l \  w_l \ {\rm e}^{+i l   {\pi\over L}   x^-}                  
  \Bigr)                                                                    
\end    {equation}                                                              
where  $ w_l  = 1/ \sqrt{l} $ and the canonical commutation
relations leading to Fock commutators
$     \bigl[ a_l , a^{\dag}_{l'} \bigr] = \delta_{l,l'} $                       
($l,l' > 0 $).                                                                  
Sometimes it will be convenient to write the Kronecker                          
$\delta_{l , l'}$ as $\delta_l^{l'} $.  

The off-diagonal components of $\Phi$                                           
are complex fields with                                                         
$ \varphi_+ (x^-) = \varphi_-^{\dag} (x^-) $. 
The momentum conjugate to $ \varphi_-$ is                                       
$ \pi^- = \bigl(\partial_- + i  g  v \bigr) \varphi_+$.
The other conjugate pair is obtained                                            
simply by hermitian conjugation.
As in \cite{PKP95,Kal96} the field expansion can be made
over half-integer momenta
\begin {equation}                                                               
  \varphi_- (x) =                                                               
    {{\rm e}^{+ i m_0  {\pi\over L}   x^-} \over \sqrt{4\pi}}\ %
  \sum_{m={1\over 2}}^{\infty} \Bigl(                                           
 \  b_m \  u_m \ {\rm e}^{-i  m   {\pi\over L}   x^-}                           
 + \  d^{\dag}_m \  v_m \ {\rm e}^{+i  m   {\pi\over L}   x^-} \Bigr)           
 \label{fockexp} 
\end {equation}                                          
where $ u_m (z) = 1/\sqrt{m + \zeta}$ and                                       
$ v_m (z) = 1/\sqrt{m - \zeta}$.                                                
The objects $m_0$ and $\zeta$ are functions of $z$,                             
defined by $m_0(z) = ({\rm{integer \, part \, of\,}} z) - \frac{1}{2},          
\zeta(z) = z - m_0(z)$. They                                                    
satisfy the relations                                                           
$ m_0 (z+1) = m_0 (z) + 1$, $m_0  (-z)= -m_0  (z)$,                     
$\zeta  (z+1) = \zeta  (z)$, and $ \zeta (-z)= -\zeta (z)$.                  
The domain interval is now $ - {1\over 2} < \zeta (z) < {1\over 2} $            
for all values of $ z $. For the fundamental domain                             
$m_0 = -\frac{1}{2}$, but the specific choice no longer matters.
The Fock modes then obey bosonic commutation relations  
$\bigl[ b_n , b^{\dag}_m \bigr] = \bigl[ d_n , d^{\dag}_m \bigr] =
\delta_{n}^{m}$ and all others zero.
Finally one notes that a large gauge transformation                             
$z\rightarrow z + 1$ produces only $m_0 \rightarrow m_0 + 1$                    
and thus only a change of the overall phase in Eq.(\ref{fockexp}).              
Most importantly                                                                
the Fock vacuum defined with respect to                                         
$b_m$ and $d_m$ is {\it invariant} under these transformations. 

The fields $A^-$, as usual in this type of gauge, are redundant
variables and are solved by implementing Gauss' law strongly. 
We do not give these equations explicitly here.
What is important is that the zero mode colour diagonal
component of the Gauss law must be imposed as a condition on
physical states. After inserting the above field expansions
this amounts to satisfying
\begin {equation}  \sum_{m = {1\over 2}}^{\infty}                                               
   \Bigl(b^{\dag}_m   b_m - d^{\dag}_m   d_m \Bigr)                             
    \vert {\rm {phys}} \rangle = 0 \ .                                          
\label{gaussop}                                                                 
\end    {equation}                                                              
Thus physical states                                                            
have equal numbers of ``b'' and ``d'' particles.

We can thus describe the complete Hilbert space in terms                        
of states with the non-separable structure: 
\begin{eqnarray}\label{genstates}                                               
|\,\Psi_i\,\rangle = \sum_{r=0}^{\infty}
\,\sum_{\nu=1}^N\, C_{r,\nu}^{(i)}      
\,\Psi_r\,|\nu\rangle                                                           
\end{eqnarray}                                                                  
with                                                                            
$C_{r,\nu}^{(i)} \not= c_r^{(i)}c_{\nu}^{(i)}$, 
$\Psi_r$ the Schr\"odinger wave function for the zero mode
and $|\nu\rangle$ a Fock space of an arbitrary number of $a$ modes
but equal number of $b$ and $d$ modes.                   

The light-cone energy, $P^-$, and momentum, $P^+$, operators                    
are obtained, as usual, from the energy-momentum tensor. 
The momentum operator turns out to be diagonal, 
$P^+= \frac{\pi}{L}\hat{K}$ with $\hat{K}$ the `harmonic resolution',  
\begin{equation}
\hat{K} =  \sum_{n=1}^\infty a_n^\dagger a_n n +                             
\sum_{m=\frac{1}{2}}^\infty \big( (m + \zeta) \ b_m^\dagger b_m                 
+ (m - \zeta) \  d_m^\dagger d_m \big)
\ . 
\end{equation}
The Hamiltonian $P^-$ is fully interacting and rather complicated.
For the present it suffices to give its form schematically.
In terms of the dimensionless operator $\hat{H}$ defined by
$ P^-= \frac{L}{\pi} \hat{H}$, we have 
\begin{equation}
\hat{H} = -\,2 \hat{g}^2\,                                                    
\frac{1}{\mbox{cos}^2(\pi\zeta)}\frac{d}{d\zeta}\,\mbox{cos}^2                  
(\pi\zeta)\frac{d}{d\zeta} \,+\, H_F
\end{equation}
where the first term is the kinetic term of the zero mode
and the second term contains the Fock operator structure,
as well as dependence on the zero mode $\zeta$ and the constrained
zero mode. As said, the constrained mode we omit. 

Our task is to solve for the low energy mass spectrum of this system
by solving the eigenvalue problem
\begin{eqnarray}                                                                
M^2 |\Psi\rangle = 2 \hat{K}\hat{H} |\Psi\rangle = 
2 K_i H_i |\Psi\rangle       
\,.                                                                             
\end{eqnarray}                                                                  
The eigenvalue $K$ of the harmonic resolution $\hat{K}$ is related to       
the size of the matrix system while  
$|\Psi\rangle $ are superpositions of states in the Hilbert
space described above. The continuum limit is then taken
by $K \rightarrow \infty$. 

In the absence of the zero mode
$\zeta$ this problem has been solved numerically now many times.
We wish now to incorporate the zero mode into these calculations.
The significant observation we make use of is that 
the vacuum of the theory, while no longer just the Fock vacuum, 
must still be a state of zero longitudinal momentum $K$ 
and can be picked out of the general Hilbert space state given above.
Indeed it turns out to be the tensor product state
$|\Omega\rangle \equiv \Psi_0[\zeta] \otimes |0\rangle$
(see also, \cite{HOT97}). 
The wavefunction $\Psi_0[\zeta]$ is an eigenstate of the quantum
mechanical problem defined by the Hamiltonian
\begin{equation}
H_0 = -\,2 \hat{g}^2\,                                                    
\frac{1}{\mbox{cos}^2(\pi\zeta)}\frac{d}{d\zeta}\,\mbox{cos}^2                  
(\pi\zeta)\frac{d}{d\zeta} \,+\,
\langle 0| H_F |0 \rangle
\ . 
\end{equation}
The Fock vacuum expectation value of the Hamiltonian thus
generates a potential governing the dynamics of the zero mode.
As everywhere in this study, the constrained mode terms here are
dropped. Formally $\langle 0| H_F |0 \rangle$ 
contains UV divergences
which can be absorbed by a mass renormalisation
 \begin{equation}                                                               
\frac{\mu^2}{4} = \frac{\mu_0^2}{4}\, + \, 2\hat{g}^2                           
\sum_{m=\frac{1}{2}}^{\Lambda}                                                  
\frac{1}{m} 
\end{equation}
with $\Lambda$ a half-integer valued large momentum cutoff. 
The same renormalisation renders the Hamiltonian finite in
the Fock sector, in the absence of the constrained zero mode.  
As shown in \cite{PKP95,Kal96} the renormalised potential
is just an infinite square-well, leading to eigenmodes
$\sqrt{2}\, \sin[\pi (r+1) (\zeta + \frac{1}{2})]$               
with $r = 0,1,2,3,...\infty$. As discussed in \cite{KPP94,PKP95},
in the continuum limit the higher modes $r \neq 0$ become
infinite in energy so that the ground state is actually
the only relevant wavefunction. We are left with  
\begin{equation}
\Psi_0 [\zeta] = \sqrt{2}\, \sin [\pi (\zeta + \frac{1}{2})]
\end{equation}
as the only relevant mode. We have thus, in an approximation
that omits the constrained field contributions \cite{Kal96},
solved the vacuum sector of the theory.

\section{Eigenvalue Problem in Presence of Zero Mode} 

Formally, the problem we are confronted with now is the
diagonalization of the following matrix 
\begin{eqnarray}\label{6}
2\,\hat{K} \int_{-\frac{1}{2}}^{\frac{1}{2}} d\zeta
  \,\Psi_0^{\dagger}(\zeta)\,\langle K;i\,|H_F|\,j;K \rangle
  \,\Psi_0 (\zeta)\,.
\end{eqnarray}
Since we are interested in the low energy states at large harmonic
resolution $K$, we can truncate the Fock space to two particle states
which certainly in the absence of zero modes is a good
approximation \cite{DKB93}. The same should hold also in
the presence of the zero mode now as the mode does not change
fundamentally the partonic structure of the theory expressed
on the light-front.    

As in \cite{PaB96}, the two-particle sector contains the lowest state 
of neutral bound scalar field pairs. 
We are able to examine two classes of orthogonal states 
\begin{eqnarray}\label{7}
\hspace{0.5cm}|aa\rangle_n \equiv 
a_n^{\dagger}a_{K-n}^{\dagger}|0\rangle 
\quad n=
1,..., \Big[\frac{K}{2}\Big] \quad\mbox{and}
\end{eqnarray}
\begin{eqnarray}\label{8}
|bd\,\rangle_m \equiv b_m^{\dagger}d_{K-m}^{\dagger}|0\rangle 
\quad m=1,..., K-\frac{1}{2}\quad.
\end{eqnarray}
The notation $[\frac{K}{2}]$ means the highest integer 
which is smaller than or equal to the numeric value $\frac{K}{2}$. 
It is necessary in order to prevent double counting 
due to the identification     
$|aa \rangle _n = | aa \rangle _{K-n}\,$ because of the commutation
relations, for all \,$n= 1,...K-1$\,.

In the two particle sector we have then  
linear combinations of the $(\frac{K}{2}\,+\,K)$--basic states 
\begin{eqnarray}
|\Psi _i\rangle = \sum _{n=1}^{[\frac{K}{2}]} C_{0,n}^{(i)} \Psi_0(\zeta)
  a_n^{\dagger}a_{K-n}^{\dagger}|0\rangle + \sum _{n=1}
^{K-\frac{1}{2}} C_{0,m}^{(i)} \Psi_0(\zeta)
  b_m^{\dagger}d_{K-m}^{\dagger}|0\rangle \quad.
\label{sampstate} 
\end{eqnarray} 
These states satisfy Gauss' law and are eigenstates of the harmonic
resolution operator, 
$ \hat{K}|aa\rangle_n = K|aa\rangle_n $ and
$\hat{K}|bd\rangle_m = K|bd\rangle_m$ 
for each $m,n$.

Taking matrix elements of $\hat{H}$ with respect to the basic
two-particle states leads to three contributions. 
Firstly, there is the pure $aa$-sector
\begin{equation}
_{n'}\langle aa |\hat{H}| aa\rangle_n 
= \Big[\frac{I_n(\zeta)}{n} + \frac{I_{K-n}(\zeta)}{K-n}\Big]
\,\Big(\delta_n^{n'}+\delta_{K-n'}^n\Big)\,,
\end{equation}
the $bd$-sector
\begin{equation}
_{m'}\langle bd |\hat{H}| bd\rangle_m 
= \Big[\frac{J_m(\zeta)}{m+\zeta} +
  \frac{J_{K-m}(-\zeta)}{K-m-\zeta}\Big]\delta_m^{m'}+S_{bd}(\zeta)
\end{equation}
and finally the mixing between these sectors 
\begin{eqnarray}
_{n'}\langle aa |\hat{H}| bd\rangle_m 
& = & S_{aa}(K-n',n';m,K-m;\zeta) + S_{aa}(n,K-n;m,K-m;\zeta), 
\nonumber \\ 
_{m'}\langle bd |\hat{H}| aa\rangle_n
&=&S_{aa}(K-n,n;m',K-m';\zeta) + S_{aa}(n, K-n ;m',K-m';\zeta).
\end{eqnarray}
The functions 
$I_n, J_n, S_{aa}$ and $S_{bd}$ have been calculated in  
\cite{Annette} and take the form 
\begin{eqnarray}
 I_n(\zeta) = \mu^2 
  &+& \frac{g^2}{16\pi} \Bigl(16 
   + 16n \sum\limits_{k=\frac{1}{2}}^{n-\frac{1}{2}}
         \frac{k^2+\zeta^2}{(k^2-\zeta^2)^2}
   + \zeta^2\sum\limits_{m=\frac{1}{2}}^\infty
       \frac{4}{m(m^2-\zeta^2)} \Bigr)\, \\
 J_m(\zeta)  = \mu^2 
   &+&  \frac{g^2}{16\pi} \biggl( 16 
    +  8(m+\zeta)\Bigl( \sum\limits_{k=\frac{1}{2}}^{m-1}
       \frac{k^2+\zeta^2}{(k^2-\zeta^2)^2}+
       \sum\limits_{k=1}^{m-\frac{1}{2}}\frac{1}{k^2}\Bigr)
   +  4\frac{m+\zeta}{(m-\zeta)^2} \biggr.
\nonumber\\
   &\phantom{+}&  \phantom{\hat g ^2 \biggl. 12\ln 2}
   + \zeta^2\sum\limits_{n=\frac{1}{2}}^\infty
       \frac{2}{n(n^2-\zeta^2)}
   + \zeta (m+\zeta)
      \sum\limits_{k=m+1}^\infty \frac{16k}{(k^2-\zeta^2)^2} \biggr)
\end{eqnarray}
and
\begin{eqnarray}
S_{aa} (n_i; \zeta) &=& 
          \left( \sqrt{\frac{n_1}{n_3+\zeta}} +
                 \sqrt{\frac{n_3+\zeta}{n_1}} \right)
          \left( \sqrt{\frac{n_2}{n_4-\zeta}} +
                 \sqrt{\frac{n_4-\zeta}{n_2}} \right)
                 \frac{(-2\hat g ^2)}{(n_1-n_3+\zeta)^2}\,,\\
S_{bd} (n_i; \zeta) &=& 
         \left( \sqrt{\frac{n_1+\zeta}{n_3+\zeta}} +
                \sqrt{\frac{n_3+\zeta}{n_1+\zeta}} \right)
         \left( \sqrt{\frac{n_2-\zeta}{n_4-\zeta}} +
                \sqrt{\frac{n_4-\zeta}{n_2-\zeta}} \right)
                         \frac{(-2)\hat g ^2}{(n_1-n_3)^2}\nonumber\\
\phantom{S_{bd} (n_i; \zeta)} &+& 
         \left( \sqrt{\frac{n_2-\zeta}{n_1+\zeta}} -
                \sqrt{\frac{n_1+\zeta}{n_2-\zeta}} \right)
         \left( \sqrt{\frac{n_4-\zeta}{n_3+\zeta}} -
                \sqrt{\frac{n_3+\zeta}{n_4-\zeta}} \right)
                \frac{2\hat g ^2}{(n_1+n_2)^2}\,.
\end{eqnarray}
In these terms then, the matrix problem in 
the two particle sector can be written 
\begin{eqnarray}
2 \hat{K}\ast
\int \limits_{-\frac{1}{2}}^{\frac{1}{2}}\, d\zeta\,
\left( \begin{array}{cc} _{n'}\langle aa |\hat{H}| aa\rangle_n
                           &_{n'}\langle aa|\hat{H}| bd\rangle_{m} 
                          \\_{m'}\langle bd |\hat{H}| aa\rangle_{n} 
                          & _{m'}\langle bd |\hat{H}| bd\rangle_m   
       \end{array}\right)* 
       \sin^2[\pi(\zeta+\frac{1}{2})]\,.
\label{2partmat} 
\end{eqnarray}

In the ideal case, one could attempt to analytically carry
out the $\zeta$ integrations in the expression of Eq.(\ref{2partmat}) and
then extract, in the continuum limit, a modified bound state 't Hooft
equation \cite{tHo74}. This has been, thus far, beyond our means.
We are content in the following with a numerical analysis of
the spectrum.   

\section {Numerical Results} 

We diagonalize the matrix Eq.(\ref{2partmat})
numerically and compare the obtained spectra 
with the spectra calculated without zero modes in \cite{PaB96}.
We stress that the spectrum without the zero modes has been
computed again here {\it independently} of \cite{PaB96}, and thus is
also a control on our computation with the zero modes.  
To make contact with the original non-dimensionally reduced                 
pure glue theory, we take the renormalised mass to be zero                      
and normalise the coupling constant $\hat{g}^2 = 1$. 
\begin{figure}
 \centerline{
\psfig{figure=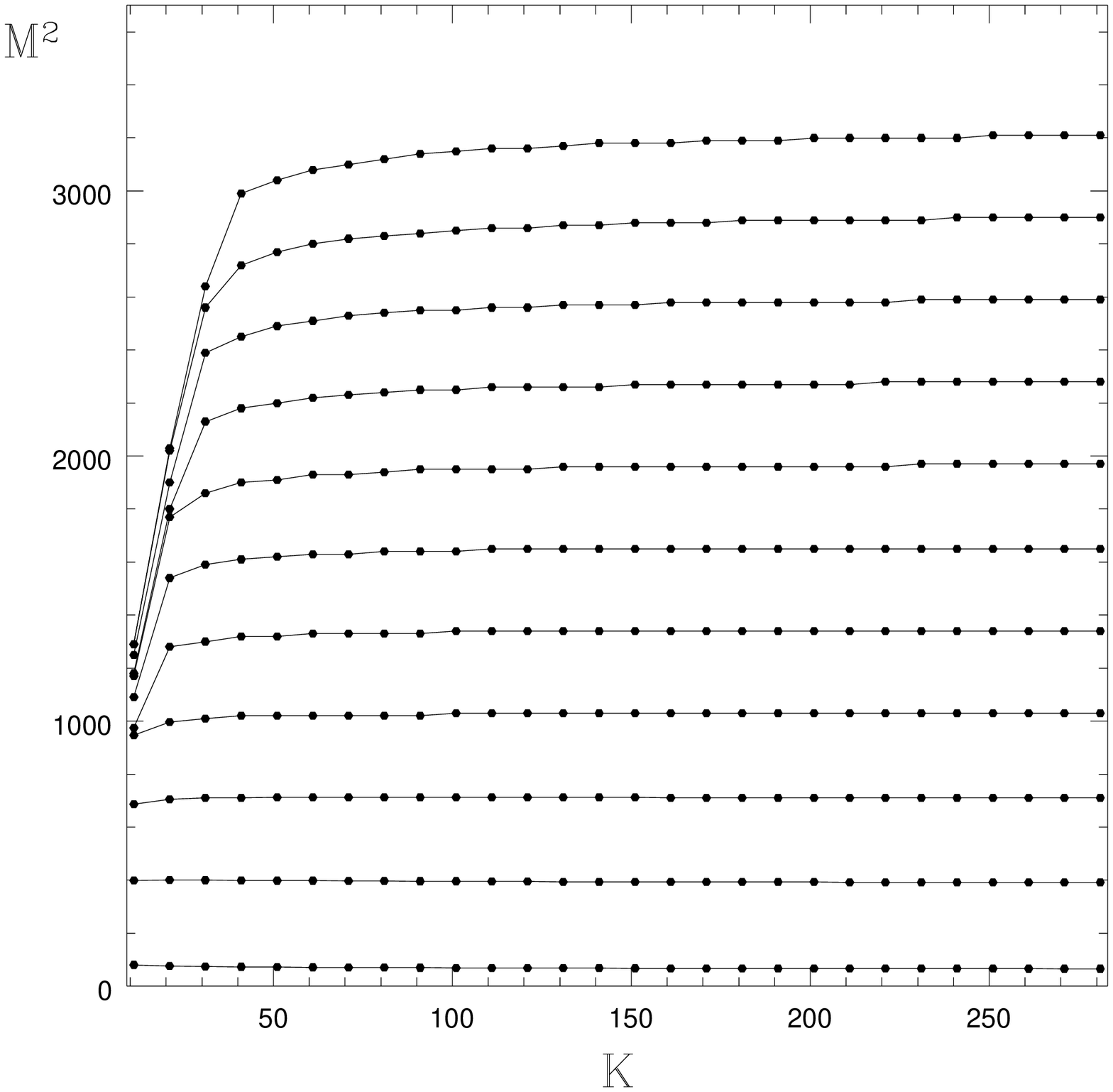,
width=8cm,angle=-0}
\psfig{figure=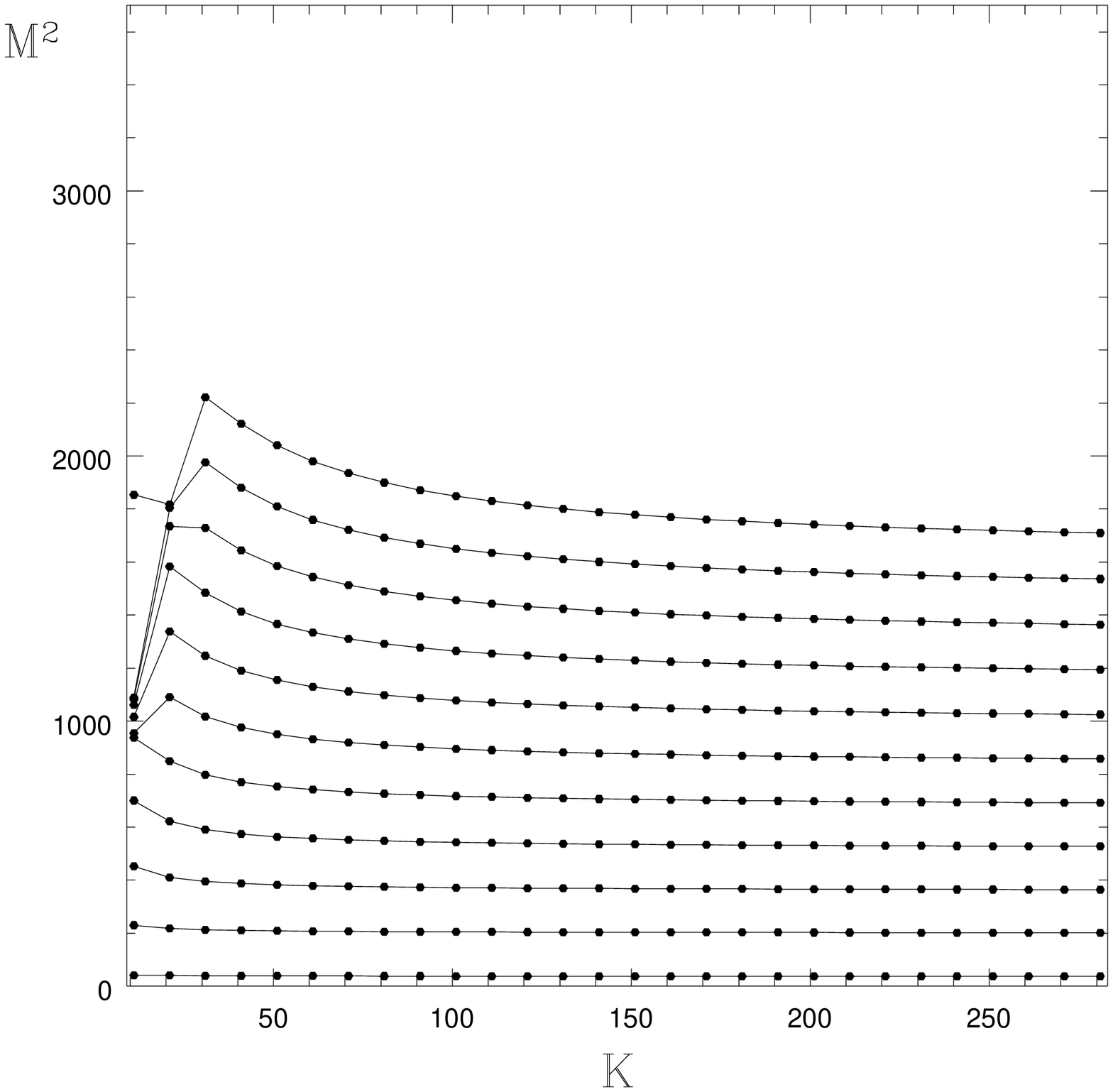,
width=8cm,angle=-0}
}
\centerline{ \parbox{5in}{ 
\caption{ \label{eigenvalues}{\sl The square of the invariant mass
    $M_i^2$ in the two particle sector as a function of the harmonic
    resolution $K$. On the left hand side the spectrum with zero mode omitted
and on the right the spectrum with the zero mode included.}}}}
\end{figure}

The result for the two spectra is shown in Fig. \ref{eigenvalues}. 
We show the first eleven eigenvalues for values of the harmonic
resolution between $K=11$ and $K=281$.
There are certainly observable differences in the two spectra,
though qualitatively they are both characterised by discrete
levels whose spacing decreases with $M^2$.  
Let us focus now on the lowest state whose invariant mass 
is most accurately described in the two particle truncation.
We are thus able to increase the harmonic resolution to
quite high values for the purpose of making a reliable extrapolation
to the continuum.
  
\begin{figure}
 \centerline{
\psfig{figure=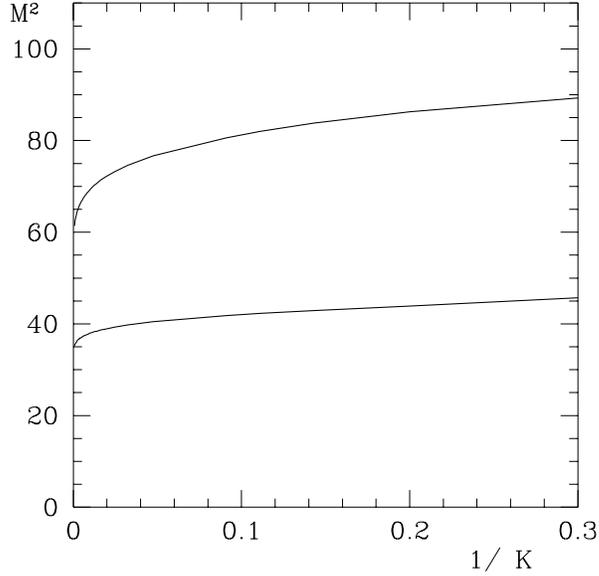
,width=8cm,angle=-0}
}
\centerline{ \parbox{5in}{ 
\caption{ \label{ground states}{\sl The ground state 
invariant mass squared including (curve above)
and omitting (curve below) the dynamical zero mode. 
The invariant mass squared is shown as a function of 
$\frac{1}{K}$.}}}}
\end{figure}

In Fig. \ref{ground states}, the ground state mass 
is shown as a function of $\frac{1}{K}$ 
for the two cases with and without the dynamical zero mode. 
We have gone up to $K=1500$. 
The mass in both cases is seen to converge 
to a finite value. 

We examine next the differences in the two values of the invariant
mass-squared as a function of $K$.
\begin{figure}
 \centerline{
\psfig{figure=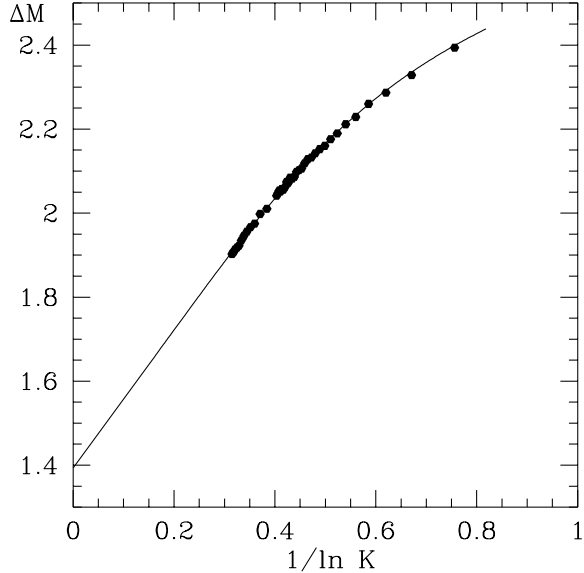,width=8cm
,angle=-0}}
\centerline{ \parbox{5in}{ 
\caption{ \label{Fit}{\sl The difference 
$\Delta M = M_{without}-M_{with}$ of the lowest eigenvalue 
without and with the dynamical zero mode. 
The dots represent the value obtained 
numerically. The curve is the fit.}}}}
\end{figure}
This is shown in Fig. \ref{Fit}. 
We numerically extrapolate to the continuum limit
by fitting the mass difference 
$\Delta M \equiv M_{without}-M_{with}$ 
to a power series in  $\frac{1}{ln\,K}$.  The best fit
turns out to be  
\begin{eqnarray}
\Delta M (K) = 1.47609 - \frac{3.36568}{(\mbox{ln\, K})} + 
\frac{5.20769}{(\mbox{ln\, K})^3} +
\frac{3.34316}{(\mbox{ln\, K})^5}
\end{eqnarray}
and gives the continuous curve in Fig.\ref{Fit}. 
To see how significant the shift is against the mass excluding the
zero mode, $\Delta M/M_{without}$, we also need the extrapolation  
\begin{eqnarray}                                                                
M_{without}(K) = 7.03883 -\frac{6.2887}{(\mbox{ln \,K})} + 
\frac{9.71745}       
{(\mbox{ln\,K})^3}                                                              
-\frac{6.5194}{(\mbox{ln\,K})^5}\quad.                                          
\end{eqnarray} 

We thus obtain for $K\rightarrow \infty$ 
\begin{eqnarray}
\frac{M_{without}-M_{with}}{M_{without}} = 0.20971
\ .
\end{eqnarray}
The extrapolation shows that the dynamical zero mode
leads to a 21 percent shift upwards in the invariant 
mass of the lowest state with respect to the mass computed with
zero modes suppressed.

\section{Summary and Outlook} 

In this work the influence of the dynamical zero mode
on the invariant mass spectrum of two-dimensional SU(2)
Yang-Mills theory has been studied. Omission of the
constrained zero mode leads to the potential for the zero
mode being an infinite square well. Taking the lowest
eigenfunction for this quantum mechanical problem, we have
been able to diagonalise the mass-squared operator
in the two particle sector. In this truncation we expect
the lowest state, if not the lowest few states,
to be accurately computable. Extrapolating numerically
to the continuum limit, we found a reduction in the mass
of the lowest state by 21\% due to the dynamics of the zero mode.

It may seem surprising that a single discrete momentum mode 
can have non-vanishing consequences in the continuum limit.
However, it should be borne in mind that this  
zero {\it momentum} mode is nonetheless an infinite number of 
{\it harmonic oscillator} modes describing the lowest
square-well eigenmode. Precisely how the Schr\"odinger
wavefunction has conspired to cause this result is not
immediately clear, though the obvious place is in the 
fact that the continuum bound-state ('t Hooft-like) equation 
becomes precisely sensitive to small momentum in the terms 
where  a principal value prescription is usually invoked
\cite{tHo74,BPR80,vdS96}. 
The most straightforward explanation of our numerical 
observation is then that the principal value prescription, 
once the zero mode wavefunctions are integrated over, 
undergoes modification.

Our result encourages further exploration within the
present two dimensional Yang-Mills model now in the presence
of the contributions of the {\it constrained} zero mode. 
As mentioned in the introduction, in \cite{Kal96} it was seen that
this mode introduces a centrifugal barrier in the
potential for the dynamical zero mode. Thus the
appropriate quantum mechanical eigenfunctions are no
longer those of the infinite square well, but rather the
double-well oscillator. Even though the size of the barrier
was found to be small in \cite{Kal96}, in light of the results of our
work, it would not be unreasonable to suspect further
modifications in the final spectrum, particularly in the mass of the
lowest state of the spectrum. Within this model then, the shift
in this lowest mass could be determined as a function 
of the potential barrier height. 
At sufficient height, the vacuum would  break 
centre symmetry spontaneously \cite{Kal96} due to mixing of 
symmetric and antisymmetric states in the double well.
It would be interesting to determine the mass of the lowest
state in the spectrum -- still a two-particle bound-state --
at this critical value of the barrier height.  

In a theory which includes fermions, alongside the
features studied in \cite{MRS97}, there will undoubtedly 
be an interplay between the specific `gluonic' structures we have 
dealt with here and states built out of quark Fock operators -- 
namely mesons and baryons.
Could this be a relevant mechanism on the light-cone 
in a realistic gauge theory for giving mass to hadronic states 
that would otherwise be thought of as massless? 
It is still far too early to tell.

\end{document}